\documentclass[12pt]{iopart}

\usepackage{graphicx}
\begin{document}

\title[Scaling Properties of Fluctuation and Correlation Results from PHENIX]{Scaling Properties of Fluctuation and Correlation Results from PHENIX}

\author{J T Mitchell for the PHENIX Collaboration \footnote{For the full list of PHENIX authors and acknowledgements, see Appendix 'Collaborations' of this volume.}}

\address{Brookhaven National Laboratory, P.O. Box 5000, Building 510C, Upton, NY 11973-5000, USA}
\ead{mitchell@bnl.gov}
\begin{abstract}
Recent surveys of multiplicity fluctuations, transverse momentum fluctuations, and two-particle azimuthal correlations are presented for several collision systems as a function of centrality and transverse momentum. Both multiplicity and transverse momentum fluctuations exhibit a power law scaling as a function of the number of participants that is independent of the collision system. Although these observations are consistent with critical behavior, the critical exponent $\eta$ measured using azimuthal correlations is seen to be independent of centrality and collision system. Also observed in the azimuthal correlations is a displaced away side peak in central Au+Au collisions when the pairs are restricted to low transverse momentum.
\end{abstract}

%\maketitle

\section{Multiplicity and Transverse Momentum Fluctuations}

The topic of event-by-event fluctuations of the inclusive charged particle multiplicity in relativistic heavy ion collisions has recently received attention due to the observation of non-monotonic behavior in the scaled variance ($\sigma^{2}/\mu$) as a function of system size at SPS energies \cite{na49MF}. PHENIX has surveyed the behavior of inclusive charged particle multiplicity fluctuations as a function of centrality and transverse momentum in $\sqrt{s_{NN}}$=62.4 GeV and 200 GeV Au+Au collisions, and in $\sqrt{s_{NN}}$=22.5, 62.4, and 200 GeV Cu+Cu collisions.

Since multiplicity fluctuations are well described by Negative Binomial Distributions (NBD) in both elementary \cite{ua5} and heavy ion collisions \cite{e802MF}, the data for a given centrality and $p_{T}$ bin are fit to an NBD from which the mean and variance are determined.  Measurements of $\sigma^{2}/\mu$ as a function of azimuthal acceptance confirm that $\sigma^{2}/\mu$ increases linearly with the azimuthal range over which the measurement is made for all species and centralities.  Hence, the results presented here are linearly extrapolated to $2\pi$ azimuthal acceptance in order to facilitate direct comparisons to other measurements. Due to the finite width of each centrality bin, there is a non-dynamic component of the observed fluctuations that is present due to fluctuations in the impact parameter within a centrality bin. The magnitude of this component is estimated using the HIJING event generator \cite{HIJING}, which well reproduces the mean multiplicity of RHIC collisions \cite{PPG019}. The estimate is performed by comparing fluctuations from simulated events with a fixed impact parameter to events with a range of impact parameters covering the width of each centrality bin, as determined from Glauber model simulations. The HIJING estimates are confirmed by comparing the HIJING fixed/ranged fluctuation ratios to measured 1\%/5\% bin width ratios. The magnitude of the scaled variance is reduced by this correction. A 15\% systematic error for this estimated is included in the error bars shown. The values of the scaled variance remain significantly above the random (Poisson) expectation of 1.0.

\begin{figure}[h]
\begin{minipage}{18pc}
\includegraphics[width=18pc]{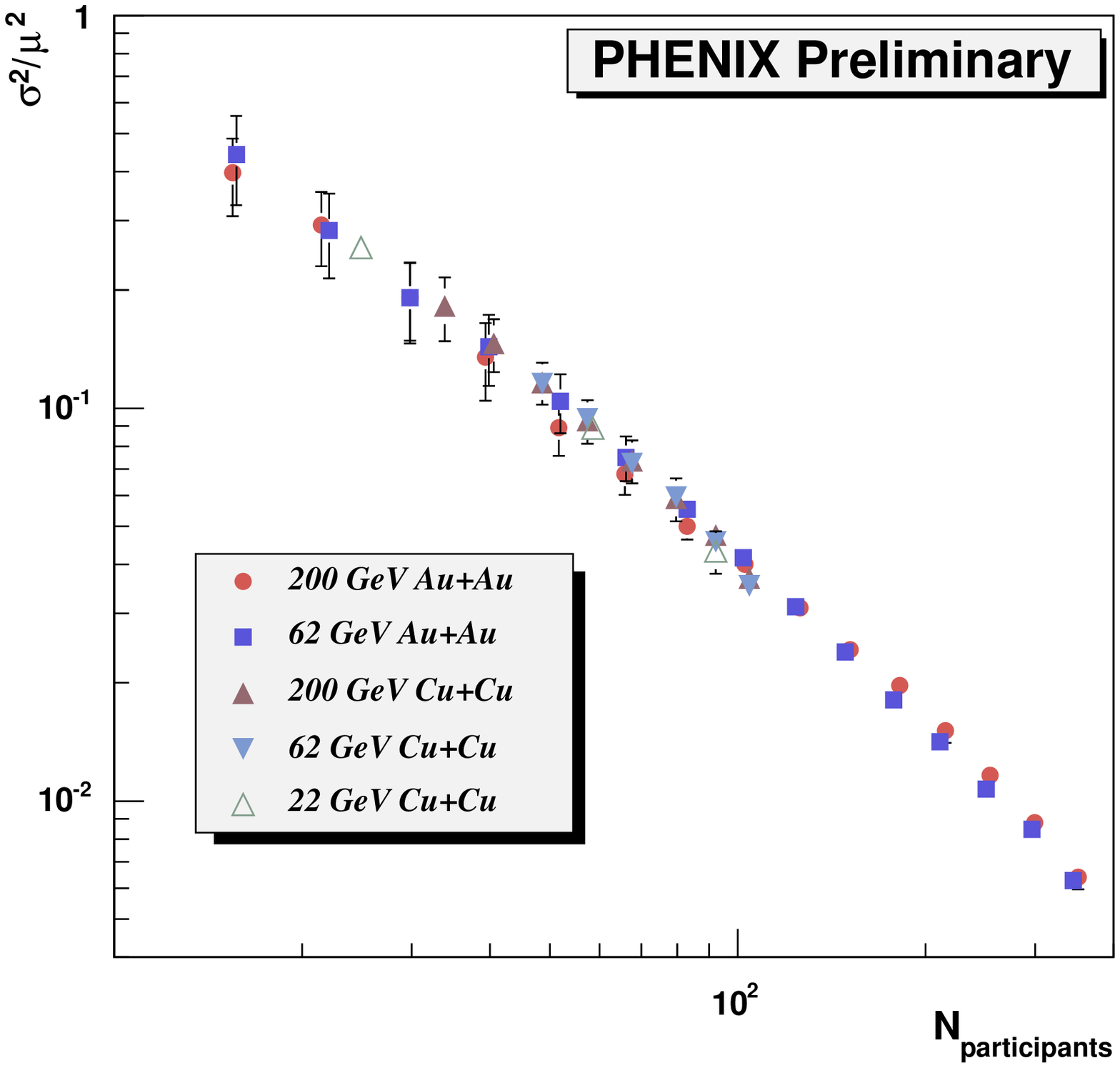}
\end{minipage}\hspace{2pc}%
\begin{minipage}{18pc}
\includegraphics[width=18pc]{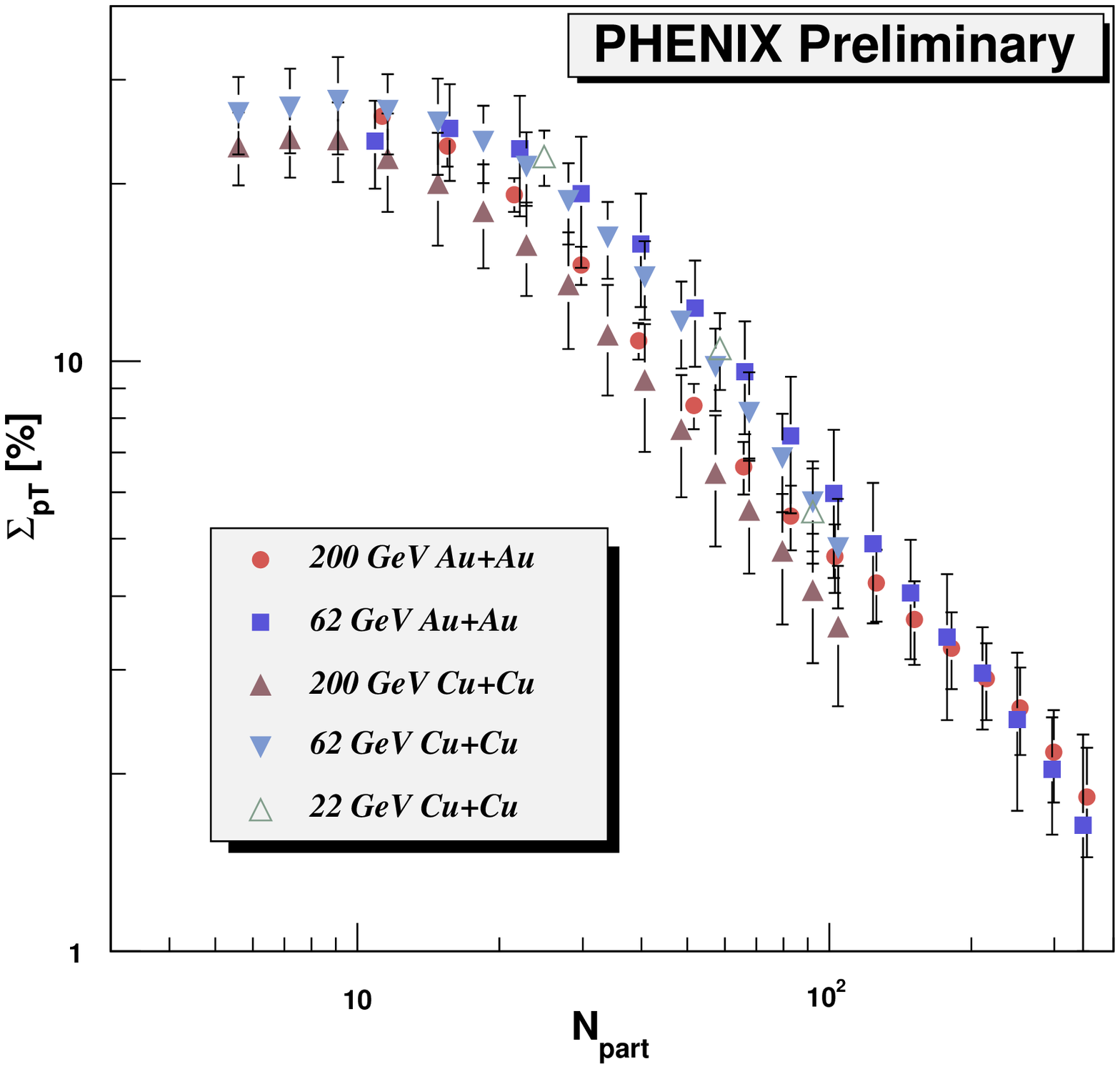}
\end{minipage}\hspace{2pc}%
\caption{\label{fig:fluc} Left: Multiplicity fluctuations for inclusive charged hadrons in the transverse momentum range $0.2<p_T<2.0$ GeV/c in terms of $\sigma^{2}/\mu^{2}$ as a function of $N_{p}$. Right: Event-by-event $p_T$ fluctuations for inclusive charged hadrons within the PHENIX acceptance in the transverse momentum range $0.2<p_T<2.0$ GeV/c in terms of $\Sigma_{p_T}$ as a function of $N_{p}$.}
\end{figure}

In the Grand Canonical Ensemble, the variance and the mean of the particle number can be directly related to the compressibility, $k_{T}$: $\sigma^{2}/\mu^{2}=k_{B}(T/V)k_{T}$, where $k_{B}$ is Boltzmann's constant, T is the system temperature, and V is its volume \cite{Stanley}. Fig. \ref{fig:fluc} (left) shows the fluctuations in terms of $\sigma^{2}/\mu^{2}$ as a function of the number of participants, $N_{p}$, for all 5 collision systems.  In order to emphasize the observed universal power law scaling property, each species has been scaled to match the 200 GeV Au+Au data. The data points for all systems can be best described by the curve $\sigma^{2}/\mu^{2} \propto N_{p}^{-1.40 \pm 0.03}$. The observed scaling is independent of the $p_T$ range over which the measurement is made.

PHENIX has also completed a survey that expands upon previous measurements of event-by-event transverse momentum fluctuations in 200 GeV Au+Au collisions \cite{ppg027}. Here, the magnitude of the $p_{T}$ fluctuations will be quoted using the variable $\Sigma_{p_T}$, as described in \cite{ceresPTF}. $\Sigma_{p_T}$ is the mean of the covariance of all particle pairs in an event, normalized by the inclusive mean $p_T$. $\Sigma_{p_T}$ is related to the inverse of the heat capacity of the system \cite{Korus}.

Fig. \ref{fig:fluc} (right) shows $\Sigma_{p_T}$ as a function of $N_{p}$ for all 5 collision systems measured over the $p_T$ range $0.2<p_T<2.0$ GeV/c. The data is shown within the effective PHENIX azimuthal acceptance of 4.24 radians. Note that the magnitude of $\Sigma_{p_T}$ exhibits little variation for the different collision energies. Since $\Sigma_{p_T}$ does not scale with the jet cross section at different energies, it is evident that hard processes are not the primary contributor to the observed fluctuations. With the exception of the very most peripheral collisions, all systems exhibit a universal power law scaling as a function of $N_{p}$. The data points for all systems are best described by the curve: $\Sigma_{p_T} \propto N_{p}^{-1.02 \pm 0.10}$. The observed scaling is independent of the $p_T$ range over which the measurement is made.

\section{Azimuthal Correlations at Low Transverse Momentum}

The study of fluctuations is important since they provide information on whether or not a system undergoes a phase transition. Correlations can also be used for this purpose. Near a critical point, several properties of a system diverge. The rate of divergence can be described by a set of critical exponents, which should be identical for any system belonging in the same universality class. The critical exponent for the correlation function is $\eta$, which can be directly extracted from the HBT component of azimuthal correlation functions: $C(\Delta \phi) \propto \Delta\phi^{-(d-2+\eta)}$, where d is the dimensionality of the system \cite{Stanley}.

\begin{figure}[h]
\begin{minipage}{18pc}
\includegraphics[width=18pc]{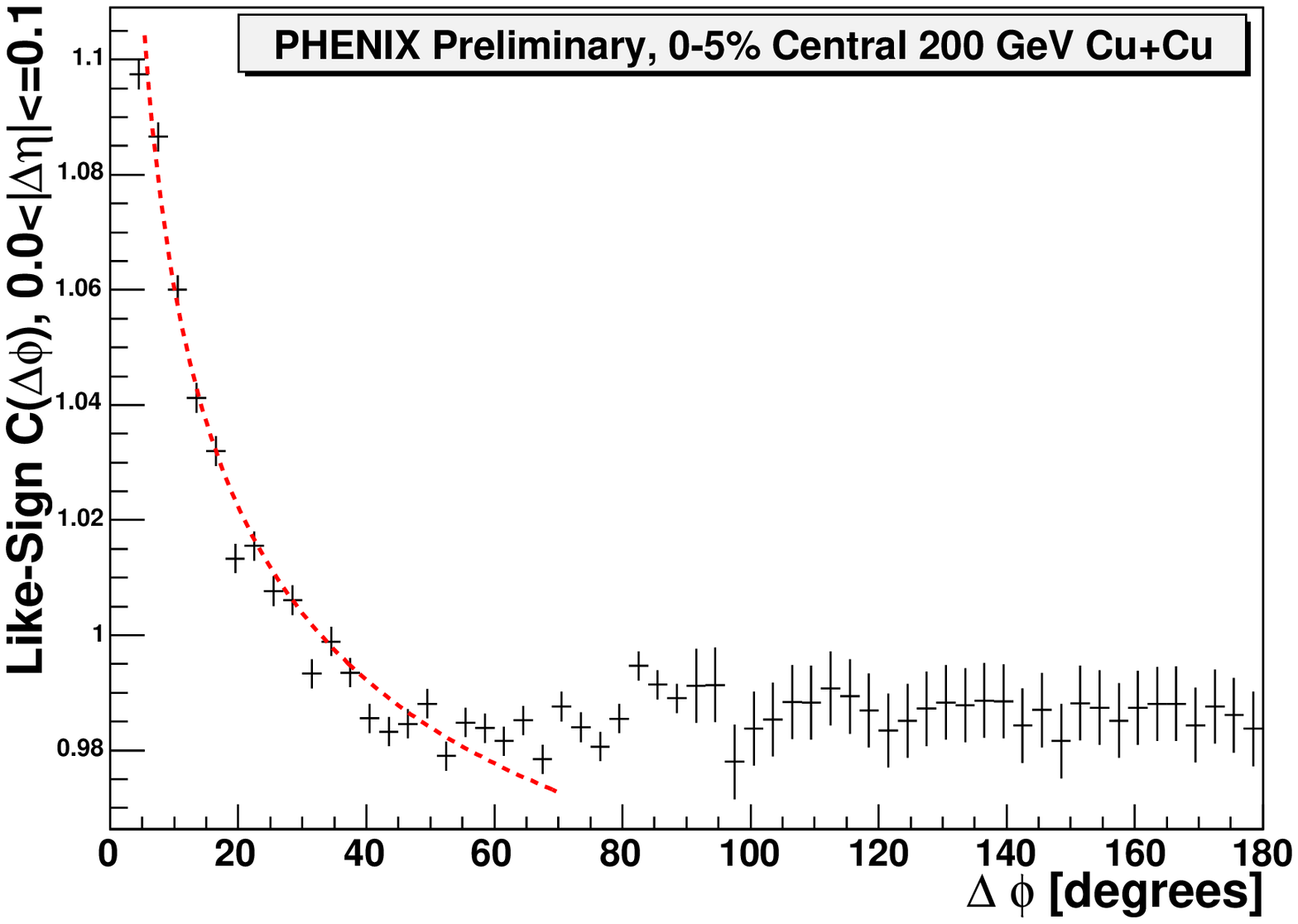}
\end{minipage}\hspace{2pc}%
\begin{minipage}{18pc}
\includegraphics[width=18pc]{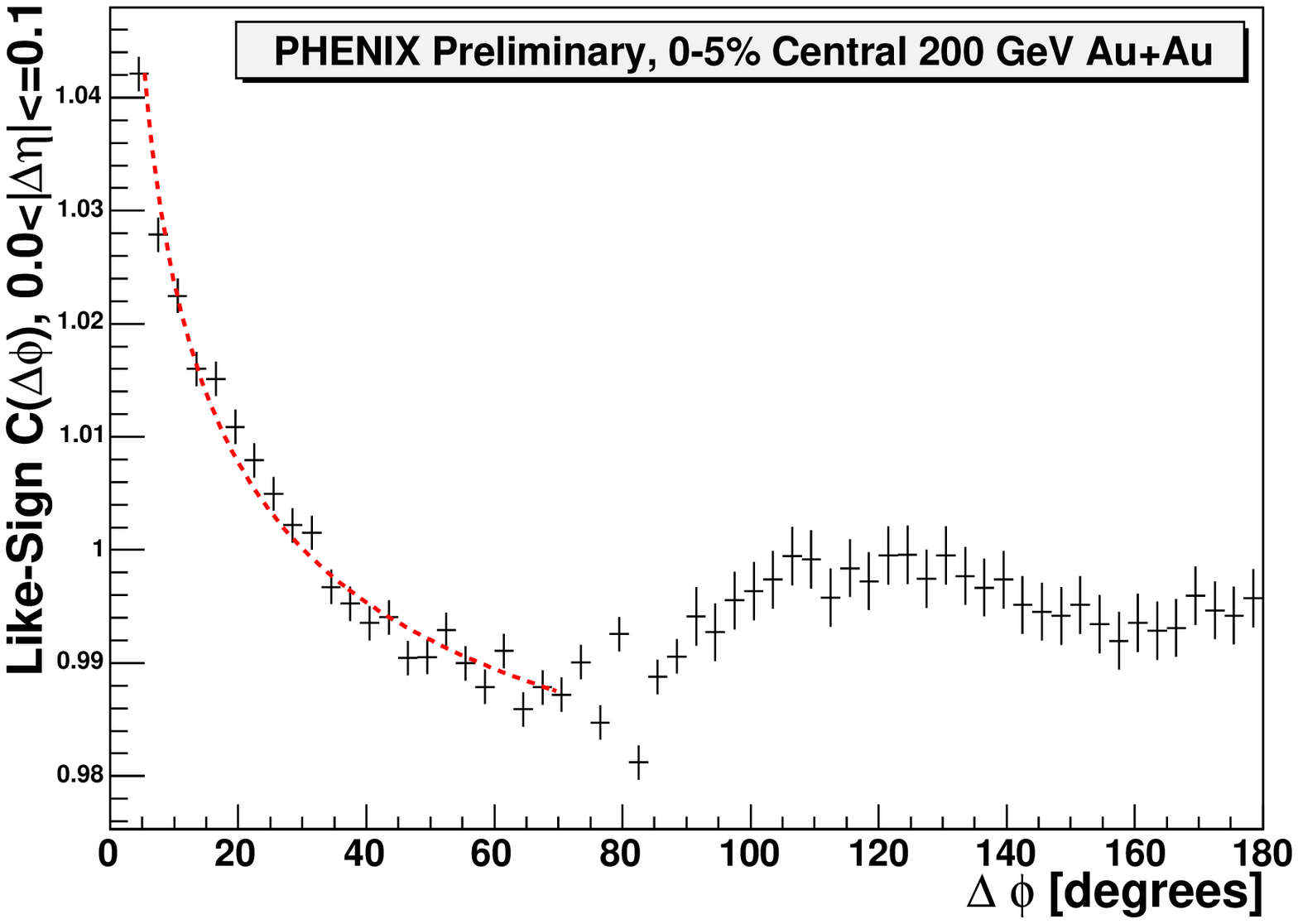}
\end{minipage}\hspace{2pc}%
\begin{minipage}{18pc}
\includegraphics[width=18pc]{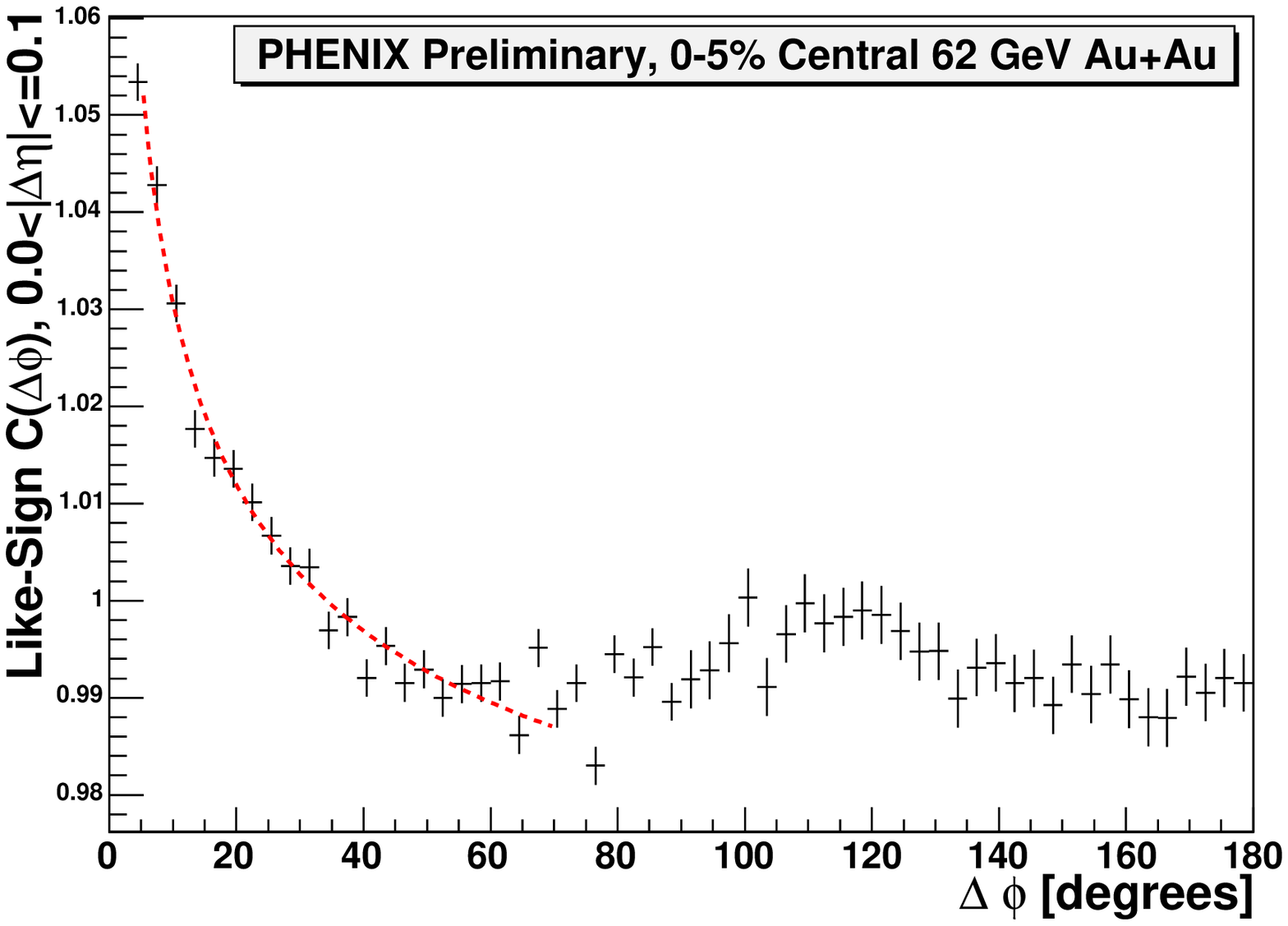}
\end{minipage}\hspace{2pc}%
\begin{minipage}{18pc}
\includegraphics[width=18pc]{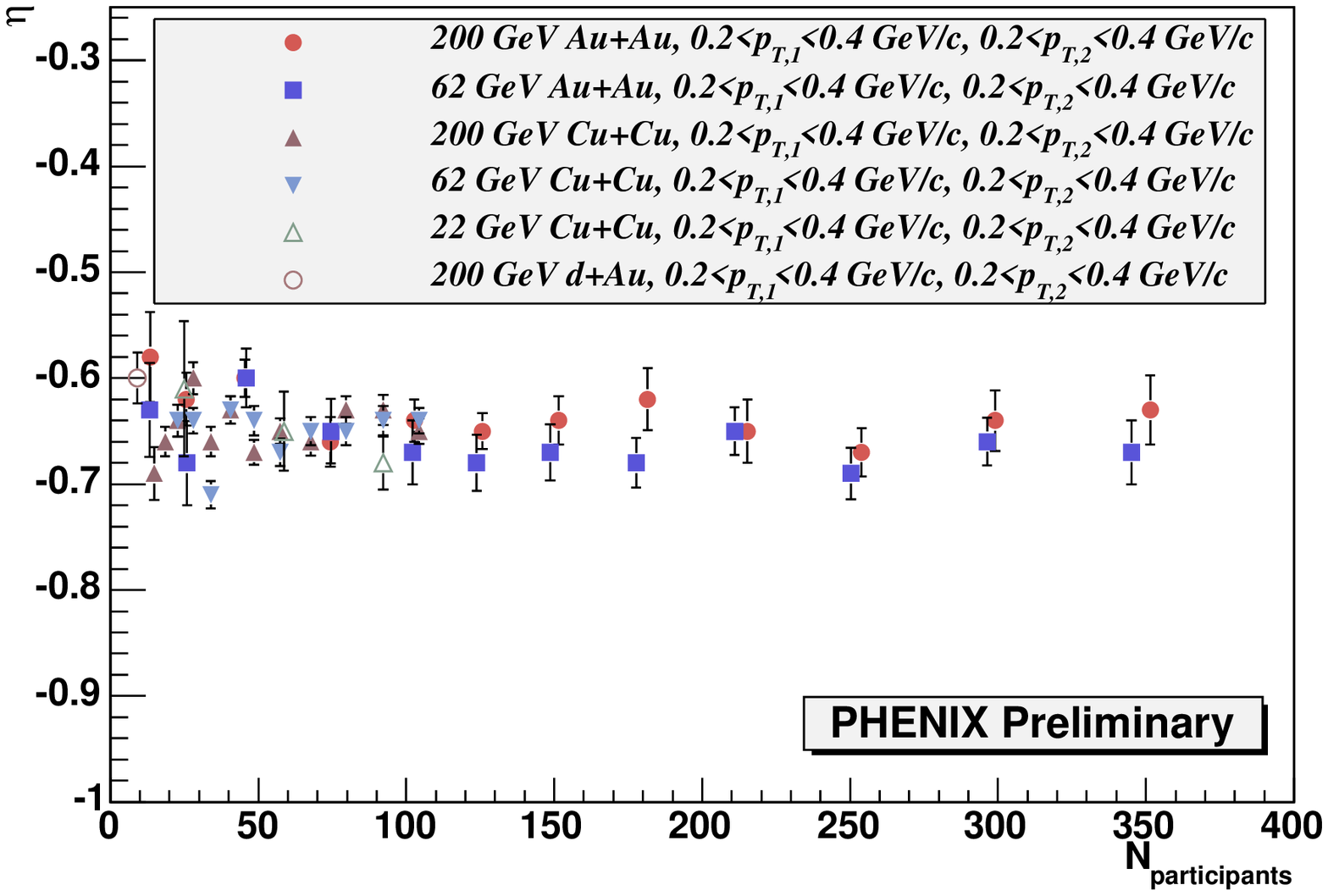}
\end{minipage}\hspace{2pc}%
\caption{\label{fig:corr} Low $p_T$ like-sign pair azimuthal correlation function for 0-5\% central 200 GeV Cu+Cu (top left), 200 GeV Au+Au (top right), and 62 GeV Au+Au (bottom left) collisions from charged hadron pairs with $0.2<p_{T,1}<0.4$ GeV/c and $0.2<p_{T,2}<0.4$ GeV/c. Bottom right: The critical exponent $\eta$ with d=3 extracted from the like-sign correlation functions as a function of $N_{p}$.}
\end{figure}

PHENIX has measured azimuthal correlation functions of like-sign pairs at low $p_T$ for several collision systems. The correlations shown will isolate the HBT peak in pseudorapidity by restricting $|\Delta\eta|<0.1$ for each particle pair. Correlations are constructed for low $p_T$ pairs by correlating all particle pairs in an event where both particles lie within the $p_T$ range $0.2<p_{T,1}<0.4$ GeV/c and $0.2<p_{T,2}<0.4$ GeV/c. Note that there is no trigger particle in this analysis. The correlation functions are constructed using mixed events as follows: $C(\Delta\phi) = \frac{dN/d\phi_{data}}{dN/d\phi_{mixed}}\frac{N_{events,mixed}}{N_{events,data}}$. Figure \ref{fig:corr} shows azimuthal correlation functions for Cu+Cu and Au+Au collisions. The dashed lines are the critical exponent function fits. Confirmation of the HBT peak has been made by observing its disappearance in unlike-sign pair correlations and by observing $Q_{invariant}$ peaks when selecting this region.

For all collision systems, including 200 GeV d+Au, the extracted value of the critical exponent $\eta$ shown in Fig. \ref{fig:corr} (bottom right) lies between -0.6 and -0.7 with d=3, independent of centrality. Since $\eta$ is constant in heavy ion collisions, does not differ from the d+Au system, and has a value that significantly differs from expectations from a QCD phase transition (e.g. if QCD belongs in the same universality class as the 3-D Ising model (d=3), $\eta$=+0.5 \cite{Reiger}), it is unlikely that critical behavior is being observed in the correlation functions measured thus far.

Measurements of several properties of the low $p_T$ correlation functions are described elsewhere \cite{jtmQM2005}. Note that a familiar feature is apparent in the low $p_T$ correlation functions for central Au+Au collisions at $\sqrt{s_{NN}}$=200 and 62 GeV. A displaced away-side peak is observed in a location that is consistent with peaks observed in correlations between particles at high $p_T$ \cite{phxHighPt}. Further study of this feature is currently underway.

\end{document}